\documentclass{aa}  

\usepackage{graphicx}
\usepackage{lipsum}
\usepackage{txfonts}
\usepackage{color}
\usepackage{amsmath,amsfonts,amssymb}
\usepackage{float}
\usepackage{hyperref}
\usepackage[version=3]{mhchem}
\usepackage{lineno}

\newcommand{\science}{Science }

\begin{document}

   \title{18-year long monitoring of the evolution of \ce{H2O} vapor in the stratosphere of Jupiter with the Odin space telescope}

   \author{B. Benmahi
          \inst{1}
          \and
          T. Cavali\'e\inst{1,2}
          \and
          M. Dobrijevic\inst{1}
          \and
          N. Biver\inst{2}
          \and
          K. Bermudez-Diaz\inst{2,3}
          \and
          Aa. Sandqvist\inst{4}
          \and
          E. Lellouch\inst{2}
          \and
          R. Moreno\inst{2}
          \and
          T. Fouchet\inst{2}
          \and
          V. Hue\inst{5}
          \and
          P. Hartogh\inst{6}
          \and
          F. Billebaud\inst{1}
          \and
          A. Lecacheux\inst{2}
          \and
          \AA. Hjalmarson\inst{7}
          \and
          U. Frisk\inst{8}
          \and
          M. Olberg\inst{9}
          \and
          The Odin Team
          }
   \institute{Laboratoire d'Astrophysique de Bordeaux, Univ. Bordeaux, CNRS, B18N, all\'ee Geoffroy Saint-Hilaire, 33615 Pessac, France\\
              \email{bilal.benmahi$@$u-bordeaux.fr}
         \and
            LESIA, Observatoire de Paris, Universit\'e PSL, CNRS, Sorbonne Universit\'e, Univ. Paris Diderot, Sorbonne Paris Cit\'e, 5 place Jules Janssen, 92195 Meudon, France
        \and
            Universit\'e Montpellier 2 Sciences et Techniques, Place E. Bataillon 30, 34095 Montpellier, France
        \and
            Stockholm Observatory, Stockholm University, AlbaNova University Center, 106 91, Stockholm, Sweden
        \and
            Southwest Research Institute, San Antonio, TX 78228, United States
        \and
            Max Planck Institut f\"ur Sonnensystemforschung, Justus-von-Liebig-Weg 3, 37077 G\"ottingen, Germany
        \and
            Department of Earth and Space Sciences, Chalmers University of Technology, Onsala Space Observatory, 439 92, Onsala, Sweden
        \and
            Omnisys Instruments AB, Solna Strandv\"ag 78, 171 54, Solna, Sweden
        \and
            Chalmers University of Technology, Gothenburg, Sweden
             }

   \date{Received 17 April 2020 ; Accepted 8 July 2020}

 
  \abstract
   {Comet Shoemaker-Levy 9 impacted Jupiter in July 1994, leaving its stratosphere with several new species, among them water vapor (\ce{H2O}).    }
   {With the aid of a photochemical model \ce{H2O} can be used as a dynamical tracer in the jovian stratosphere. In this paper, we aim at constraining vertical eddy diffusion ($K_{zz}$) at the levels where \ce{H2O} resides.   }
   {We monitored the \ce{H2O} disk-averaged emission at 556.936\,GHz with the Odin space telescope between 2002 and 2019, covering nearly two decades. We analyzed the data with a combination of 1D photochemical and radiative transfer models to constrain vertical eddy diffusion in the stratosphere of Jupiter.   }
   {The Odin observations show us that the emission of \ce{H2O} has an almost linear decrease of about 40\% between 2002 and 2019. We can only reproduce our time series if we increase the magnitude of $K_{zz}$ in the pressure range where \ce{H2O} diffuses downward from 2002 to 2019, i.e. from $\sim0.2$\,mbar to $\sim$5\,mbar. However, this modified $K_{zz}$ is incompatible with hydrocarbon observations. We find that, even if allowance is made for the initially large abundances of H$_2$O and CO at the impact latitudes, the photochemical conversion of H$_2$O to CO$_2$ is not sufficient to explain the progressive decline of the H$_2$O line emission, suggestive of additional loss mechanisms.    }
   {The $K_{zz}$ we derived from the Odin observations of \ce{H2O} can only be viewed as an upper limit in the $\sim0.2$\,mbar to $\sim$5\,mbar pressure range. The incompatibility between the interpretations made from \ce{H2O} and hydrocarbon observations  probably results from 1D modeling limitations. Meridional variability of \ce{H2O}, most probably at auroral latitudes, would need to be assessed and compared with that of hydrocarbons to quantify the role of auroral chemistry in the temporal evolution of the \ce{H2O} abundance since the SL9 impacts. Modeling the temporal evolution of SL9 species with a 2D model would be the next natural step.   }


   \titlerunning{Jupiter's stratospheric vertical eddy mixing constrained from \ce{H2O} long-term monitoring with Odin}
   \authorrunning{B. Benmahi et al.}
   \maketitle
%
\section{Introduction}
From the first observations of water (\ce{H2O}) in the stratospheres of giant planets \citep{Feuchtgruber1997}, the existence of external sources of material to these planets, such as rings, icy satellites, interplanetary dust particles (IDP), and cometary impacts, was demonstrated. Indeed, \ce{H2O} cannot be transported from the tropospheres to the stratospheres due to a cold trap at the tropopause of all these planets. Regarding the nature of the external sources, it is now demonstrated that Enceladus plays a major role in delivering \ce{H2O} to Saturn's stratosphere \citep{Waite2006,Hansen2006,Porco2006,Hartogh2011,Cavalie2019}, while an ancient comet impact is the favored hypothesis in the case of Neptune for carbon monoxide (CO), hydrogen cyanide (HCN) and carbon monosulfide (CS) \citep{Lellouch2005,Lellouch2010,Hesman2007,Luszcz-Cook2013,Moreno2017}. At Uranus, the situation remains unclear \citep{Cavalie2014,Moses2017}.

In July 1994, astronomers witnessed the first extraterrestrial comet impact when the Shoemaker-Levy 9 comet hit Jupiter. Several fragment impacts were observed around $-44^{\circ}$ latitude \citep{Schultz1995, Sault1997, Griffith2004}, which delivered several new species, including \ce{H2O} \citep{Lellouch1995, Bjoraker1996}. Piecing together several observations of \ce{H2O} vapor in the infrared and submillimeter with the Infrared Space Observatory (ISO), the Submillimeter Wave Astronomy Satellite (SWAS), Odin and Herschel, it was established that Jupiter's stratospheric \ce{H2O} comes from the SL9 comet impacts \citep{Bergin2000,Lellouch2002,Cavalie2008,Cavalie2012,Cavalie2013}.

\citet{Cavalie2012} used the monitoring of the \ce{H2O} emissions to try and constrain the vertical eddy mixing in Jupiter's stratosphere. Their sample of Odin observations only covered 2002 to 2009 and did not allow them to unambiguously demonstrate that the line emission was decreasing with time, as was expected from the comet impact scenario. Fortunately, the Odin space telescope is still in operation and has continued ever since to regularly monitor the \ce{H2O} emission from the stratosphere of Jupiter. In this paper, we extend the monitoring presented in \citet{Cavalie2012} by adding new observations from 2010 to 2019, hence doubling the time baseline. While \ce{H2O} is not as chemically stable as e.g. HCN \citep{Moreno2006,Cavalie2013} and can, in principle, not be used to constrain horizontal diffusion without a robust chemistry and diffusion model, we assume oxygen chemistry is now sufficiently well-known after recent progress \citep{Dobrijevic2014,Dobrijevic2016,Dobrijevic2020,Loison2017} and use \ce{H2O} nonetheless as a tracer to constrain vertical diffusion in Jupiter's stratosphere, similarly to HCN, CO and carbon dioxide (\ce{CO2}) in \citet{Moreno2003}, \citet{Griffith2004}, and \citet{Lellouch2002,Lellouch2006}. Our work therefore assumes \ce{H2O} has small meridional variability by the time of our first observation in 2002, i.e. of the order of that measured by \citet{Moreno2007} in HCN and \citet{Cavalie2013} in \ce{H2O} (a factor of 2-3). With nearly two decades of data, we can probe the layers from the level where \ce{H2O} was originally deposited by the comet to its current location by following its downward diffusion with our spectroscopic observations.

We present the Odin observations made between 2002 and 2019 in Section \ref{section:obs}. We introduce the photochemical and radiative transfer models that we used in this study in Section \ref{section:models}. Results of both the photochemical model and the analysis of Odin observations are given in Section \ref{section:results}, followed by a discussion in Section \ref{section:discussion} on the eddy diffusion profile. We give our conclusion in Section \ref{section:conclusion}.


\section{Observations}
\label{section:obs}

Odin \citep{Nordh2003} is a Swedish-led space telescope of 1.1m in diameter. It was launched into polar orbit in 2001, at an altitude of 600 km. It observes in the submillimeter domain in the frequency bands of 486-504 GHz and 541-581\,GHz. The observations of the \ce{H2O} (1$_{10}$-1$_{01}$) line at 556.936\,GHz in Jupiter's stratosphere used in this paper were made with the Submillimeter and Millimeter Radiometer \citep{Frisk2003} and the Acousto-Optical Spectrometer \citep{Lecacheux1998} using the Dicke switching observation mode. This mode is the standard Odin observation mode \citep{Olberg2003,Hjalmarson2003}. It enables integrating on a target and a reference position on the sky by using a Dicke mirror. This enables compensating for short-term gain fluctuations. In addition, a few orbits are integrated on the sky 15’ away from the source to remove other effects not corrected by the Dicke switching technique, like ripple continuum and continuum spillover from the main beam.

\begin{figure}[ht]
    \centering
    \includegraphics[width=9cm, keepaspectratio]{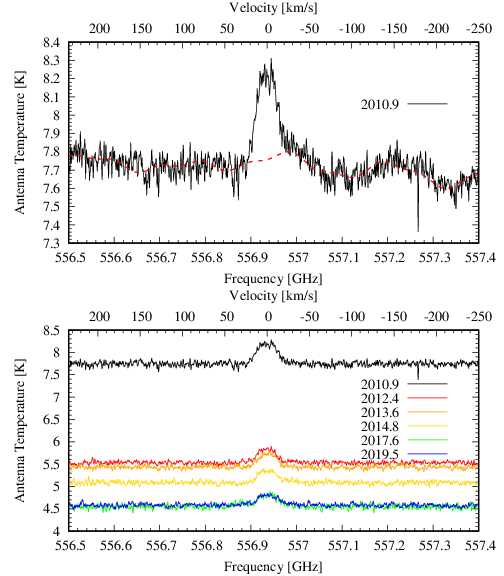}
    \caption{(Top) Example of the 2010 raw observation of H$_2$O in the stratosphere of Jupiter (black solid line) and continuum baseline (red dashed line) removed by using a Lomb periodogram. This observation was the most affected by continuum ripples. Later spectra barely showed any continuum baselines and are thus not shown here for clarity.  (Bottom) Baseline-subtracted spectra recorded after 2010 and used in this study in complement to the observations of \citet{Cavalie2008,Cavalie2012}.}
    \label{fig:rawobs}
\end{figure}

A first monitoring of Jupiter's stratospheric \ce{H2O} emission at 557GHz was already carried out by Odin over the 2002-2009 period \citep{Cavalie2012}. We have obtained additional data between 2010 and 2019, on the following dates: 2010/11/20, 2012/02/17, 2012/02/24, 2012/10/05, 2013/03/01, 2013/10/04, 2013/10/27, 2014/04/04, 2014/10/17, 2015/04/19, 2016/12/16, 2018/02/02, 2019/02/22, and 2019/10/09. We thus double the time coverage of the Odin monitoring. For each observation date, we have accumulated on average 9 orbits of integration time, for the 9 orbits we have 6 ($\times\sim$1h integration) ON Jupiter and 3 OFF at 15' to remove the residual background that we get in the Dicke switch scheme. Each observation was reduced with the same method as in \citet{Biver2005} and \citet{Cavalie2012}. Residual continuum baselines were removed using a normalized Lomb periodogram (see Fig. \ref{fig:rawobs} top) to produce the baseline-subtracted spectra we analyzed in what follows (see Fig. \ref{fig:rawobs} bottom).

Odin's primary beam is about 126\arcsec at 557\,GHz, whereas the apparent size of Jupiter is about 35\arcsec~as Odin observes when Jupiter is in quadrature. We have thus obtained disk-averaged spectra. Even though the temporal evolution of the disk-averaged \ce{H2O} vertical distribution following the SL9 impacts implies that two different dates should correspond to two different vertical profiles, we chose when possible to average the observations by groups of two or three not too far apart in time to increase the signal-to-noise ratio (S/N). All 2012 observations have been averaged into a single observation and we link this observation to an equivalent date of 2012/05/21 for our modeling. We have proceeded similarly with all 2013 observations (equivalent date of 2013/09/08), with the 2014 and 2015 observations (equivalent date of 2014/10/29), with the 2016 and 2018 observations (equivalent date of 2017/07/22), and with the 2019 observations (equivalent date of 2019/06/17). With the initial 2010 and final 2019 observations, we have a total of 6 new spectra covering 2010-2019. To further increase the S/N, we smoothed the spectra from their native spectral resolution of 1.1\,MHz to 10\,MHz. This has a very limited impact on the line shape, given that the line is already substantially smeared by the rapid rotation of the planet (12.5\,km/s at the equator). 

\begin{figure*}[ht]
    \centering
    \includegraphics[width=17cm, keepaspectratio]{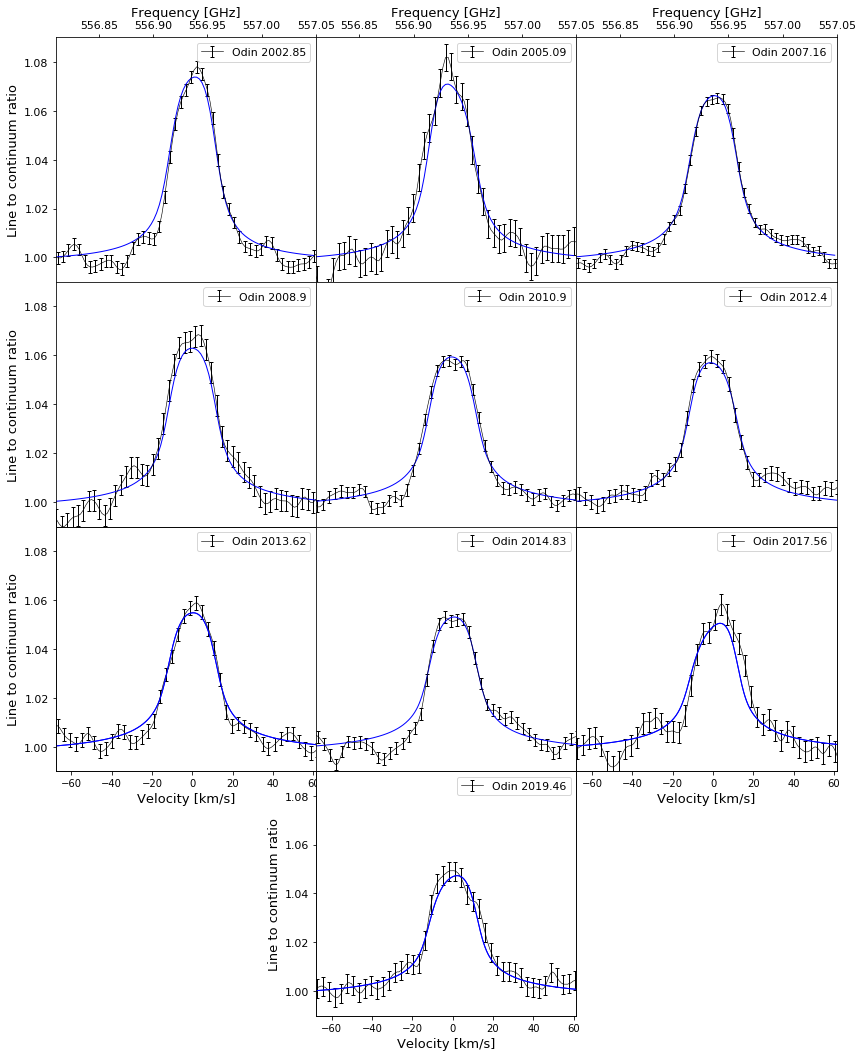}
    \caption{Odin observations of \ce{H2O} at 556.936\,GHz in Jupiter's stratosphere between 2002 and 2019. The spectral resolution is 10\,MHz. The blue lines correspond to our nominal temporal evolution model (obtained with $K_{zz}$ Model B).
    }
    \label{fig:obs_10}
\end{figure*}

\begin{figure}[h]
  \centering
   \includegraphics[width=9cm, keepaspectratio]{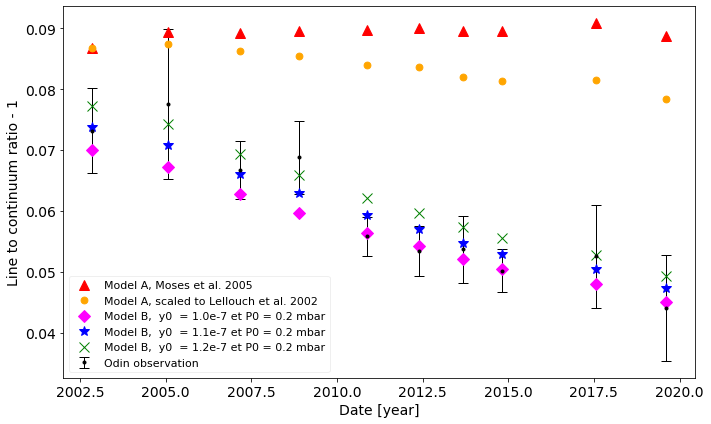}
      \caption{Evolution of the \ce{H2O} line-to-continuum ratio observed by Odin in the atmosphere of Jupiter (black points). The blue stars represent the results obtained with our $K_{zz}$ Model B. The green and pink dots correspond to different values of the $y_{0}$ parameter with the same model. The red triangles stand for our nominal parameters of $y_{0}$ and $p_{0}$ and $K_{zz}$ Model A (profile from \citealt{Moses2005}). The orange dots depict the results obtained with the vertical profiles of Model A after rescaling their respective column densities to the temporal evolution modeled by \citet{Lellouch2002} with their chemistry-2D transport model.
              }
         \label{fig:evo_amplitude}
\end{figure}

The 10 spectra that span the 2002-2019 time period and that we used in our analysis are shown in Fig. \ref{fig:obs_10}. 
Given the limited sensitivity per spectral channel of our observations, there is very limited vertical information that can be directly retrieved from the line profile. The main information then resides in the line area. In addition, the line width is mainly controlled by the rapid rotation of the planet, so that the line amplitude (l) remains the only diagnostic for temporal variability. Because observations were carried out at different Odin-Jupiter distances, there is non-negligible variability in the beam filling factor. To get rid of this variability and only keep the variability caused by the evolution of the water abundance, we divided the spectra by their observed antenna temperature continuum (c) to produce and subsequently analyze line-to-continuum ratio (l/c) spectra.
We computed the l/c by averaging the peak of the line over a range of $\pm$5\,km/s and the continuum excluding the central $\pm$50\,km/s. It has the benefit of cancelling out the variable beam dilution effect that results from the variable Jupiter-Odin distance from one another date and that impacts similarly the observed line amplitude and continuum. The evolution of the l/c of the Odin observations between 2002 and 2019 is presented in Fig. \ref{fig:evo_amplitude}. We note that long-term stability of Odin's hot calibrator is better than 2\% and is accounted for in the total power calibration scheme. It has no effect on the temporal evolution of the l/c. In addition, any detector sensitivity changes over the course of this monitoring would have similar effects on both continuum and line amplitude. So the temporal evolution seen on the l/c in Fig. \ref{fig:evo_amplitude} is only caused by changes in the H$_2$O abundance.

\section{Models}
\label{section:models}

In this section, we present the models used to reproduce the decrease in the \ce{H2O} l/c at 557 GHz observed by the Odin space telescope between 2002 and 2019. These calculations were carried out with a 1D time-dependent photochemical model to simulate the \ce{H2O} disk-averaged mole fraction vertical profile in the atmosphere of Jupiter after the SL9 impact at each observation date, and a radiative transfer code to simulate the Odin spectra. We first present the photochemical model, then the radiative transfer model, and finally our modeling strategy.

\subsection{Photochemical model}
The 1D time-dependent photochemical model used in the present study is adapted from the recent model developed for Neptune by \citet{Dobrijevic2020}, which couples ion and neutral hydrocarbon and oxygen species. The ion-neutral chemical scheme remains unchanged (see \citealt{Dobrijevic2020} for details). In the following sections, we only outline the parameters specific to Jupiter used in this model.

\subsubsection{Boundary conditions \label{sec:boundary}}
In the first step of the 1D photochemical modeling, we assumed a background flux of \ce{H2O}, CO and \ce{CO2} supplied by a constant flux of IDP with influx rates $\Phi_i$ at the top boundary given by \citet{Moses2017}: $\Phi_{\mathrm{H}_2\mathrm{O}} = 4\times 10^4$ cm$^{-2}$s$^{-1}$, $\Phi_{\mathrm{CO}} = 175\times \Phi_{\mathrm{H}_2\mathrm{O}}$, $\Phi_{\mathrm{CO}_2} = 2.5\times \Phi_{\mathrm{H}_2\mathrm{O}}$. We also account for the internal source of CO with a tropospheric mole fraction of 1\,ppm \citep{Bezard2002}. Unlike previous photochemical models, we did not include a downward flux of atomic hydrogen at the upper boundary to account for additional photochemical production of H in the higher atmosphere. We assumed that photo-ionization and subsequent ionic chemistry were responsible for this source previously added to the models. All other species were assumed to have zero-flux boundary conditions at the top of the model atmosphere (corresponding to a pressure of about $10^{-6}$ mbar).

At the lower boundary (1 bar), we set the mole fraction of He, \ce{CH4} and \ce{H2} respectively to $y_\mathrm{He} = 0.136$, $y_{\mathrm{CH}_4} = 1.81\times 10^{-3}$ and $y_{\mathrm{H}_2} = 1.0 - y_\mathrm{He} - y_{\mathrm{CH}_4}$ (see \citealt{Hue2018} for details). All other compounds have a downward flux given by the maximum diffusion velocity $v = K_{zz}/H$ where $K_{zz}$ is the eddy diffusion coefficient and $H$ the atmospheric scale height at the lower boundary.

\subsubsection{Temperature and vertical transport}
\begin{figure}[h]
   \centering
   \includegraphics[width=9cm, keepaspectratio]{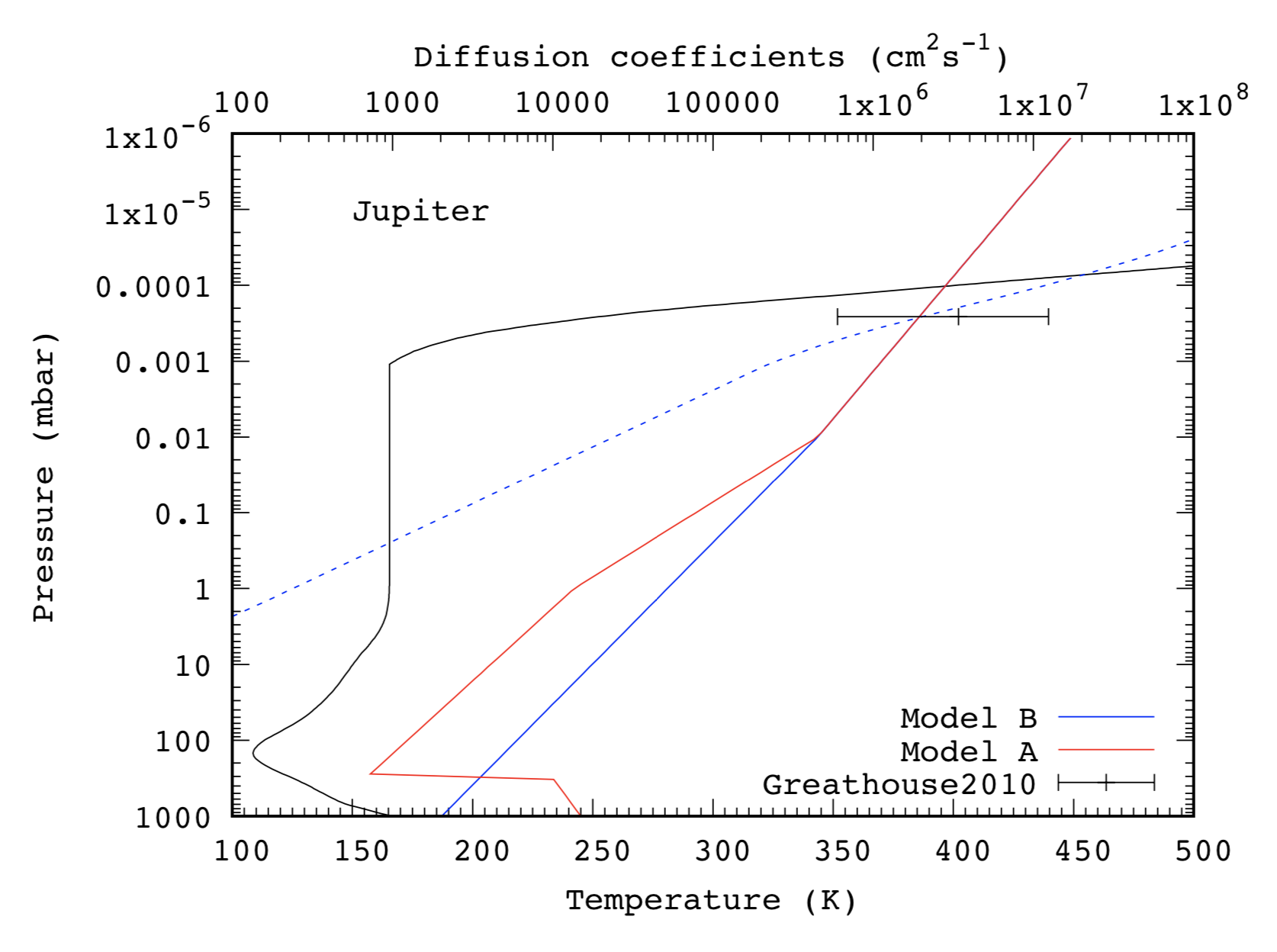}
      \caption{Temperature-pressure profile taken from \citet{Hue2018} (black solid line). Eddy diffusion coefficient $K_{zz}$ profile from \citet{Moses2005} (red solid line - Model A) compared to our nominal profile (blue solid line - Model B). The \ce{CH4} homopause occurs where the $K_{zz}$ profile crosses the \ce{CH4} molecular diffusion coefficient profile (blue dashed lines). The $K_{zz}$ value derived by \citet{Greathouse2010} at this level is shown for comparison. Our nominal $K_{zz}$ is unconstrained from our \ce{H2O} observations for pressures higher than $\sim$5\,mbar.}
         \label{fig:kzz}
   \end{figure}

The pressure-temperature profile used in the present study for all observation dates is shown in Fig. \ref{fig:kzz}. Details on this profile can be found in \citet{Hue2018}. We chose to use this disk-averaged temperature profile throughout the 18-year observation period of Odin, because Jupiter barely shows disk-averaged seasonal variability \citep{Hue2018}. In addition, \citet{Cavalie2012} already showed that reasonable disk-averaged stratospheric temperature variations could not explain the \ce{H2O} l/c evolution in the 2002-2009 period. Since the l/c decrease has continued ever since, the disk-averaged stratospheric temperature would have had to drop continuously by $\geq$10\,K over the 2002-2019 period. Even though such variability can be seen locally, such disk-averaged variability is contradicted by observations \citep{Greathouse2016}.

Our baseline $K_{zz}$ eddy diffusion profile (Model A in what follows) is the Model C of \citet{Moses2005}. This profile ensures having a \ce{CH4} mole fraction profile in agreement with observations of \citet{Greathouse2010} around the homopause. To fit the temporal evolution of the \ce{H2O} emission seen by Odin, we had to adjust this profile in the pressure range probed by the \ce{H2O} line. More details are given in Section \ref{section:procedure}. The resulting eddy profile (Model B hereafter) shown in Fig. \ref{fig:kzz} and has a simple expression given by:
$K_{zz} = K_\mathrm{ref}*(p_\mathrm{ref}/p(z))^a$
where $K_\mathrm{ref} = 4.5\times 10^5$ cm$^2$ s$^{-1}$, $p_\mathrm{ref} = 10^{-2}$ mbar and $a = 0.4$ if $p(z)<p_\mathrm{ref}$ and $a = 0.469$ otherwise.


\subsection{Radiative transfer model}
We applied the radiative transfer model described in \citet{Cavalie2008a,Cavalie2019} and used the temperature profile as well as the output mole fraction profiles of the photochemical model. Both are therefore applied uniformly in latitude and longitude over the jovian disk.

Details regarding Jovian continuum opacity, spectroscopic data and the effect of the jovian rapid rotation can be found in \citet{Cavalie2008}. We adopt the following broadening parameters (pressure-broadening coefficient $\gamma$ and its temperature dependence $n$) for \ce{H2O}, \ce{NH3} and \ce{PH3}: $\gamma_{\mathrm{H}_2\mathrm{O}}=0.080$\,cm$^{-1}$atm$^{-1}$, $n_{\mathrm{H}_2\mathrm{O}}=0.60$, $\gamma_{\mathrm{NH}_3}=0.072$\,cm$^{-1}$atm$^{-1}$, $n_{\mathrm{NH}_3}=0.73$, $\gamma_{\mathrm{PH}_3}=0.100$\,cm$^{-1}$atm$^{-1}$, and $n_{\mathrm{PH}_3}=0.70$ \citep{Dick2009,Fletcher2007,Levy1993,Levy1994}. The final spectra are smoothed to a resolution of 10\,MHz.

Odin's pointing has been checked twice a year and has remained stable within a few arcsec since its launch. However, larger pointing errors can occur when Odin is pointing to Jupiter close to occultation by the Earth (i.e. at the beginning and at the end of observations during an orbit). Only one star-tracker can then be used for the platform pointing stability, the other one pointing at the Earth. It results in a significant decrease of the pointing performance. For each Odin observation, we therefore used radiative transfer simulations to fit any east-west pointing error. Despite the large Odin beam-size with respect to the size of Jupiter, an east-west pointing error will slightly modify the weights of the various emission regions of the rotating planet during the beam convolution, and blue-shift the line if there is an eastward pointing error, or red-shift the line if there is an westward pointing error. The offsets we found are: $+$15.6\arcsec~for 2002.86, $-$34\arcsec~for 2005.09, $+$8\arcsec~for 2007.16, $-$6\arcsec~for 2008.90, $-$5\arcsec~for 2010.90, $-$16\arcsec~for 2012.40, $+$5\arcsec~for 2013.62, $+$8\arcsec~for 2014.83, $+$51\arcsec~for 2017.56, and $+$20\arcsec~for 2019.46. The vast majority of values remain smaller than one fifth of the beam size. The 2017.56 spectrum, which has the largest pointing error, is also unsurprisingly the one with the lowest quality fit. The pointing offset also marginally affects the l/c and can be best seen in the results obtained using the \citet{Moses2005} eddy mixing profile (red triangle in Fig. \ref{fig:evo_amplitude}), where some small jumps in the l/c are present (e.g. compare the 2017 point to the surrounding 2014 and 2019 ones). The l/c is also affected in principle by north-south pointing errors, especially if the thermal field is not meridionally uniform. However, we have no means to constrain such error.


\subsection{Modeling procedure} 
\label{section:procedure}
The photochemical model was used in two subsequent steps, for a given $K_{zz}$ profile. First, we ran our model with the background oxygen flux until the steady state was reached for all the species. The results of this steady state\footnote{A model with only neutral chemistry was also run to study the effect of the ionic chemistry on the photochemistry of Jupiter and to confirm what \citet{Dobrijevic2020} found for Neptune. Indeed, the ion-neutral coupling affects the production of many species in Neptune's atmosphere. In particular, it increases the production of aromatics and strongly affects the chemistry of oxygen species. We find similar effects in Jupiter.} then served as a baseline for the second step of the modeling. 

In this second step, we treated the cometary impact in a classical way \citep{Moreno2003,Cavalie2008}. We have considered a sporadic cometary supply of \ce{H2O} in July 1994 with two parameters: the initial mole fraction of \ce{H2O} $y_{0}$ deposited above a pressure level $p_0$. This level was measured by e.g. \citet{Moreno2003} and found to be 0.2$\pm$0.1\,mbar. We thus fixed $p_0$ to 0.2\,mbar in our study. The value of $y_0$ was then found by chi-square minimization and usually found close to the values reported by \citet{Cavalie2008,Cavalie2012}. We also added a CO component with a constant mole fraction of 2.5$\times$10$^{-6}$ for $p<p_0$ at the start of our simulations, in agreement with \citet{Bezard2002} and \citet{Lellouch2002}, to account for \ce{H2O}-CO chemistry. The model was then run for integration times corresponding to the time intervals between the comet impacts and the Odin observation dates. The abundance profiles were extracted for each Odin observation date. We then simulated the \ce{H2O} line at 556.936\,GHz line for each date and compared the resulting spectra with the observations by using the $\chi^2$ method. 

We started with $K_{zz}$ Model A, and adjusted it subsequently to obtain Model B by cycling the whole procedure until a good fit of all the \ce{H2O} lines was obtained.


\section{Results}
\label{section:results}

Fig. \ref{fig:evo_amplitude} shows that the decrease of the l/c, only hinted in the first half of the monitoring \citep{Cavalie2012}, is now demonstrated, with a decrease of $\sim$40\%~between 2002 and 2019. This is evidence that the vertical profile of \ce{H2O} has evolved within this time range and we thus used it to constrain vertical transport from our modeling.

We first estimated the level of residence of \ce{H2O} as a function of time with forward radiative transfer simulations using parametrized vertical profiles in which the \ce{H2O} mole fraction is set constant above a cut-off pressure level. Despite the limited S/N of our observations, we were able to estimate these levels as a function of time. The most noticeable result is that we see the downward diffusion of \ce{H2O} as the cut-off level evolves from $\sim$0.2\,mbar to $\sim$5\,mbar over the 2002-2019 monitoring period. This is the pressure range in which we could constrain $K_{zz}$.

For each $K_{zz}$ profile we tested, we explored a range of $y_0$ values (with $p_0$ always fixed to 0.2\,mbar) and generated the \ce{H2O} vertical profile for each Odin observation between 2002 and 2019. We then compared the lines resulting from these profiles with the observations in terms of l/c, and searched for acceptable fits (using a reduced $\chi^2$ test) of the temporal evolution of the \ce{H2O} l/c at 557\,GHz. The best-fit value of $y_{0}$ were usually found close to the values of \citet{Cavalie2008,Cavalie2012}, which were in agreement with previous ISO observations \citep{Lellouch2002}.

\begin{figure}[h]
   \centering
   \includegraphics[width=9cm, keepaspectratio]{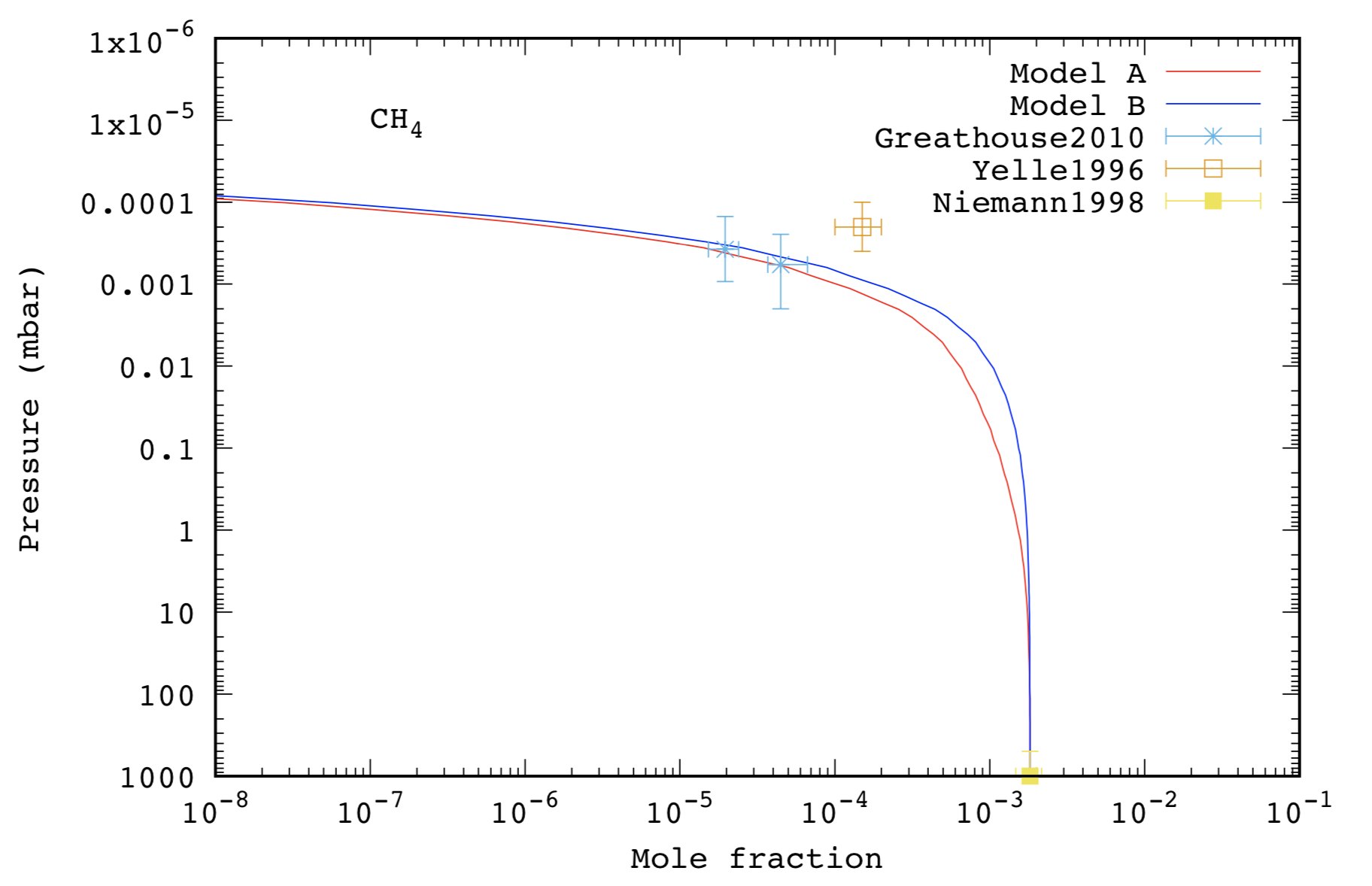}
   \includegraphics[width=9cm, keepaspectratio]{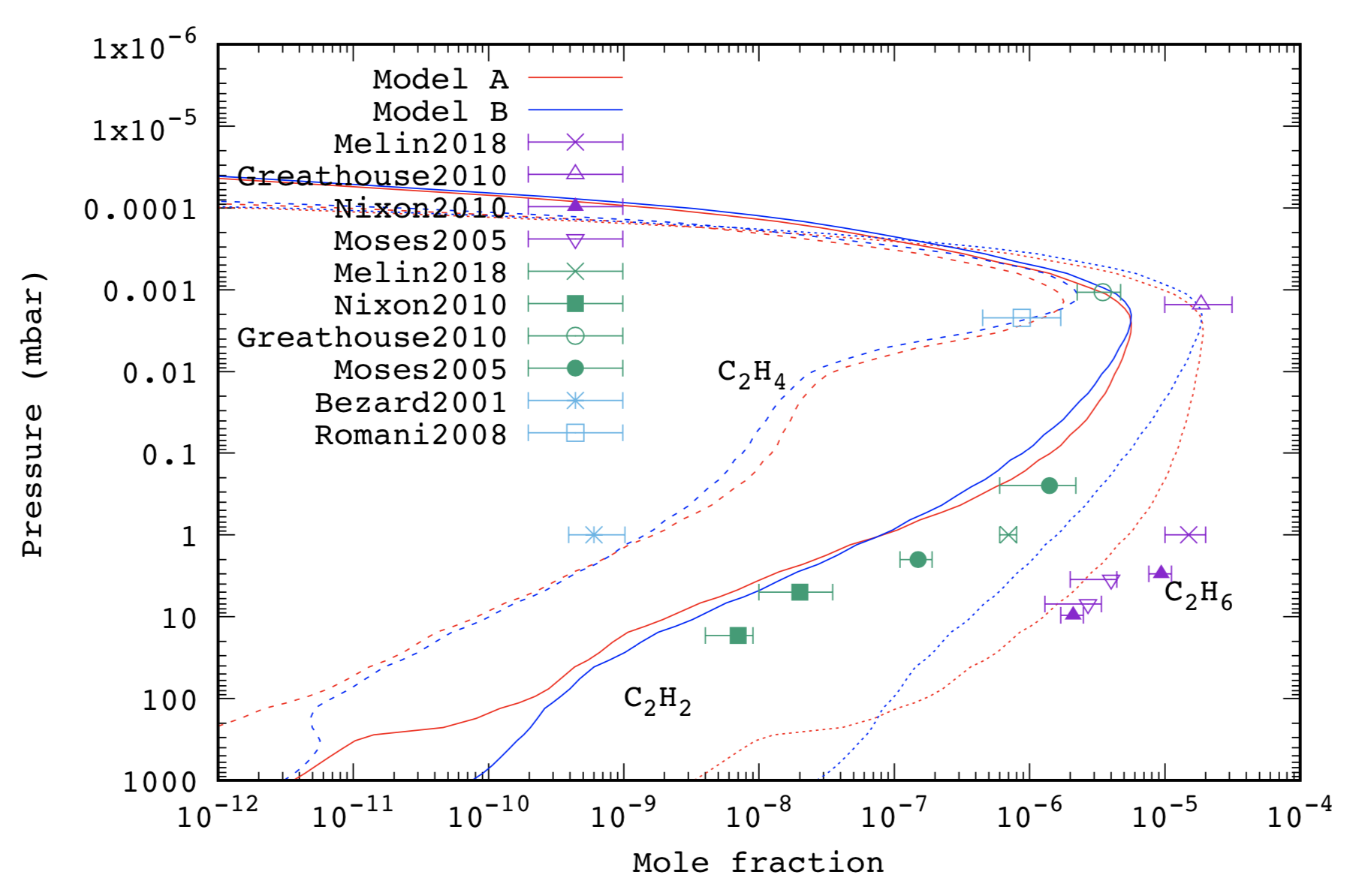}
      \caption{(Top) Mole fraction profiles of \ce{CH4} using the $K_{zz}$ Models A (red line) and B (blue line). Observations of \citet{Greathouse2010} are given for comparison. 
      (Bottom) Mole fraction profiles of \ce{C2H2}, \ce{C2H4}and \ce{C2H6} (same layout). Several observations are given for comparison (see \citealt{Moses2017} for details).}
         \label{fig:hydrocarbons}
\end{figure}

\begin{figure}[!ht]
   \centering
   \includegraphics[width=9cm, keepaspectratio]{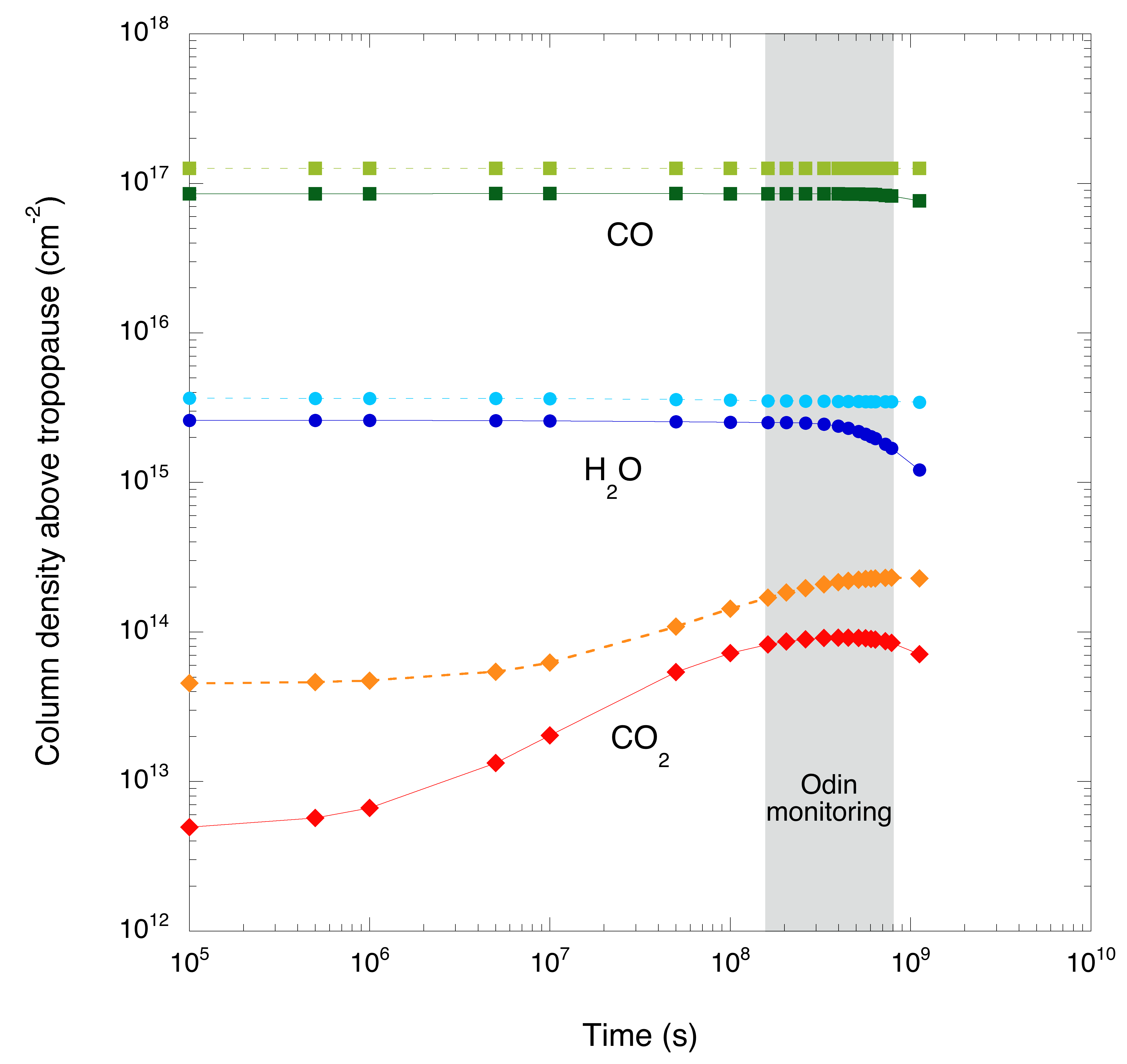}
      \caption{Column densities as a function of time after the SL9 impacts (1994) for CO, \ce{H2O}, and \ce{CO2}. Model A results in the light green, cyan and orange curves, while Model B results in the dark green, dark blue and red ones. The offset between the two models results from the different background column resulting from the IDP source (see Section \ref{sec:boundary}). The period covering the Odin monitoring (2002-2019) is highlighted in grey.}
         \label{fig:column_density}
\end{figure}

We first noted that there is no ($y_0$,$p_0$) combination that enables fitting the Odin l/c for all dates when using the $K_{zz}$ profile derived by \citet{Moses2005} (our Model A), even though main hydrocarbon observations\footnote{Results for the other species are not depicted in the paper, but can be obtained upon request.} are reproduced (see Fig. \ref{fig:hydrocarbons}). The red points in Fig. \ref{fig:evo_amplitude} show, for instance, the results obtained using our nominal parameters for $y_{0}$ and $p_{0}$ (see below). In the first years after the impacts, we find that a small fraction of \ce{H2O} (and CO) is converted into \ce{CO2}, as shown by \citet{Lellouch2002} and also previously found by \citet{Moses1996} for impact sites. The main difference between the two studies is that our model is a 1D, globally-averaged, model with complete and up-to-date chemistry, while in their study of the evolution of \ce{H2O} and \ce{CO2}, \citet{Lellouch2002} used a simplified chemistry model but a latitude-dependent model describing the spatial evolution of the SL9-generated compounds due to meridional eddy mixing. The initial disk-averaged \ce{H2O} and CO abundances are thus lower in our Model A than in those in the narrow latitudinal band in \citet{Lellouch2002}. This is also  true for our Model B (see hereafter). Given our assumed initial CO and \ce{H2O} values, we find a loss of 5\% of the \ce{H2O} column between the impacts and 2019, and only 1\% between 2002 and 2019 (see Fig. \ref{fig:column_density}), does not translate into a proportional decrease of the spectral line l/c.

We note, however, that the loss in the first years following the impacts is likely to be underestimated in this model (and so would the production of \ce{CO2}), because the \ce{H2O} and CO abundances were $\sim$10 times higher in the latitudes around the impacts \citep{Lellouch2002}. A simple 1D simulation with such abundances (Model A', initial CO mole fraction of 2.5$\times$10$^{-5}$ above 0.2\,mbar, i.e. 10 times more than in Model A) until 1997 (i.e., the date of the ISO observations of \citealt{Lellouch2002}) leads to a loss of 31\% of the \ce{H2O} column (vs. only 3\% in Model A). However, even if the initial loss is indeed underestimated in our Model A, the slope of the l/c between 2002 and 2019 would not be significantly altered between Model A and a model that would start with the conditions of Model A' and continue with disk-averaged abundances at the start of our Odin campaign (still assuming that the factor of 2-3 horizontal variability seen in \ce{H2O} in 2009 by \citealt{Cavalie2013} is small enough that it can be neglected). The slope might even be flatter given that more \ce{H2O} would have been consumed in the first place. After 2002, the loss of \ce{H2O} would then be even slower. Actually, if we take the vertical profiles of Model A and scale their respective column densities to match the temporal evolution of the chemistry-2D transport model of \citet{Lellouch2002} (from their Fig. 12), we find an intermediate case shown in Fig. \ref{fig:evo_amplitude}. However, this model falls short by a factor of 4 to reproduce the temporal decrease of the l/c observed between 2002 and 2019, as it only produces a decrease of the l/c of $\sim$10\%.

\begin{figure*}[!ht]
   \centering
   \includegraphics[width=15cm, keepaspectratio]{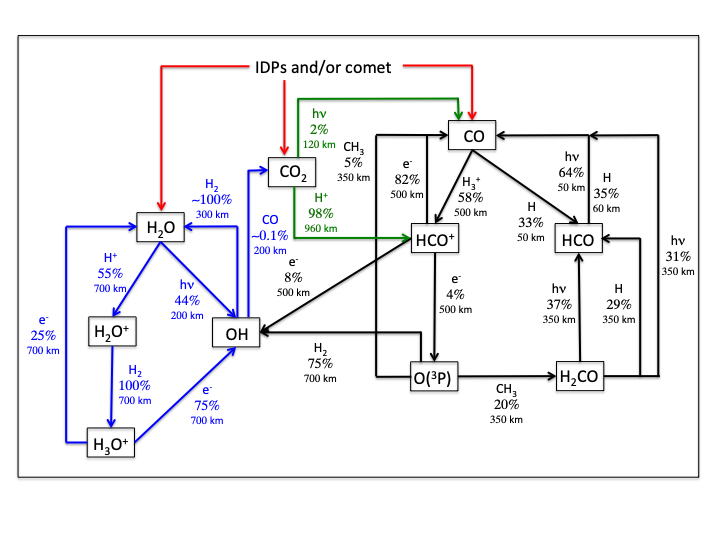}
      \caption{Simple schematic of the chemical network of oxygen species based on integrated chemical loss term over altitude. This illustrates the fate of oxygen species from the external input (IDPs and/or comet) of \ce{H2O}, CO and \ce{CO2} in the atmosphere of Jupiter. For each species, the main loss processes are given. Photolysis is represented by h$\nu$ and for reactions, the other reactant is given as a label. The percentage of the total integrated loss term over altitude is given with the altitude at which this process is maximum. Blue: sub-chemical scheme of \ce{H2O}-related species. Black: sub-chemical scheme of CO-related species. Green: sub-chemical scheme of \ce{CO2}-related species. Percentages change slightly depending on the amount of water in the atmosphere (i.e. before and after the comet impact), but the whole scheme stays the same.
Values given here correspond to the state of the chemistry just after the influx of \ce{H2O} due to the impact.}
         \label{fig:diagram}
\end{figure*}

\begin{figure}[h]
    \centering
    \includegraphics[width=9cm, keepaspectratio]{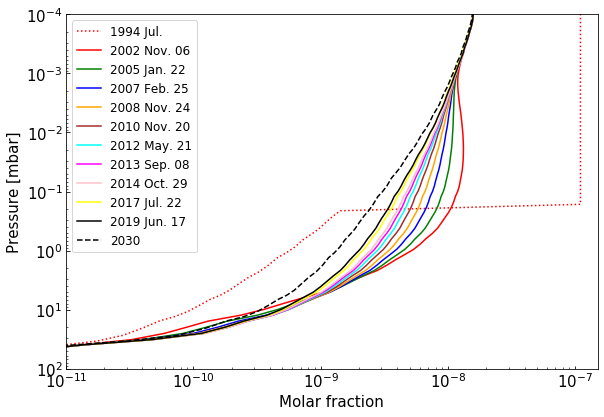}
    \caption{Evolution of the \ce{H2O} abundance profile in the stratosphere of Jupiter for $y_{0} = 1.1\times 10^{-7}$ above a pressure level $p_{0} = 0.2$ mbar (SL9 parameters) and $K_{zz}$ Model B. These profiles are obtained from the photochemical model and a comparison with the observations. The red dotted abundance profile represents the initial profile of \ce{H2O} at the time of the SL9 comet impacts in 1994. Each solid curve represents the abundance profile of \ce{H2O} at dates corresponding to Odin observations. The black dashed profile represents the abundance of \ce{H2O} that we predict for 2030.}
    \label{fig:DIFF}
\end{figure}

By increasing the magnitude of $K_{zz}$ in the millibar and submillibar pressure ranges, for instance from 1.4$\times$10$^4$cm$^2$s$^{-1}$ to 5.2$\times$10$^4$cm$^2$s$^{-1}$ at 1\,mbar and from 7.8$\times$10$^4$cm$^2$s$^{-1}$ to 1.5$\times$10$^5$cm$^2$s$^{-1}$ at 0.1\,mbar, we could accelerate the decrease of the l/c to before the start of the Odin monitoring. With this $K_{zz}$ Model B, shown in Fig. \ref{fig:kzz}, we were able to fit the pattern of the l/c temporal evolution within error bars. Fig. \ref{fig:evo_amplitude} shows our best results (green crosses, blue squares, and pink stars). These results show that the initial disk-averaged \ce{H2O} mole fraction deposited by SL9 above the 0.2 mbar pressure level was likely in the range [$y_{0} = 1.0\times 10^{-7}$, $y_{0} = 1.2\times 10^{-7}$]. In Model B, we find that the column of CO$_{2}$ tops at 0.1\% of the total O column (which again must be underestimated) and starts decreasing after 2$\times$10$^8$\,s ($\sim$6 years after the impacts), when the production of \ce{CO2} by the CO$+$OH reaction becomes less efficient than \ce{CO2} photolysis. The production and loss mechanisms for oxygen species as of 2019 are summarized in Fig. \ref{fig:diagram}. It essentially shows that \ce{H2O} is efficiently recycled and is only lost due to condensation. \ce{CO2} is lost to CO and \ce{H2O} via photolysis. The evolution of the \ce{H2O} abundance profile according to our Model B (with $y_{0} = 1.1\times 10^{-7}$ above a pressure level $p_{0} = 0.2$ mbar) is shown in Fig. \ref{fig:DIFF} at the time of the SL9 impacts and of each Odin observation. We also show a prediction for 2030 when JUICE (Jupiter Icy Moons Explorer) will start observing Jupiter's atmosphere. We note that the simulation gives us a decrease of the \ce{H2O} abundance as a function of time for pressures lower than $\sim$5\,mbar between 2002 and 2019, while it tends to increase at higher pressures because of vertical mixing.

We finally verified the agreement of Model B at the steady state for the main hydrocarbons. Fig. \ref{fig:hydrocarbons} (top) shows that our \ce{CH4} profile remains in good agreement with the \citet{Greathouse2010} observations, which is not surprising since model A and B share a common homopause. However, the \ce{C2H6} profile (Fig. \ref{fig:hydrocarbons} bottom) is incompatible with the observations, questioning the validity of $K_{zz}$ Model B. This profile is not a unique solution and properly deriving the error bars on its vertical profile would require a full retrieval, which is beyond the scope of this paper. However, we performed several tests to look for other ($K_{zz}$, $y_0$, $p_0$) combinations to explain the observed l/c evolution and found our $K_{zz}$ Model B to be quite robust as there is no solution that enables fitting both the H$_2$O temporal evolution and the hydrocarbon vertical profiles simultaneously. There is thus a contradiction between the Odin \ce{H2O} monitoring observations and the hydrocarbon observations in terms of vertical mixing, when interpreted with a 1D time-dependent photochemical model.


\section{Discussion}
\label{section:discussion}

In a 1D photochemical model, vertical transport is dominated by molecular diffusion at altitudes above the homopause and eddy mixing below this limit. The $K_{zz}$ profile is a free parameter of 1D photochemical models. The best way to constrain this parameter is to compare the model results with observational data of some particular species. In the case of Titan, an inert species like argon (Ar) is very useful for this purpose. For the giant planets, the situation is more difficult. The homopause can be constrained using \ce{CH4} observations in the upper atmosphere since its profile is driven by molecular diffusion. Below the homopause, the $K_{zz}$ profile is usually constrained from a comparison between observations and model results for the main hydrocarbons (\ce{C2H2} and \ce{C2H6}). Unfortunately, this is an imprecise methodology since model results have strong uncertainties (see for instance \citealt{Dobrijevic2010} for Neptune and \citealt{Dobrijevic2011} for Saturn), which can be much larger than uncertainties on observational data. In a recent photochemical model of Titan, \citet{Loison2019} used \ce{H2O} and HCN to constrain $K_{zz}$ in the lower stratosphere. One of the reasons is that the chemical processes that drive their abundances are expected to be simpler than for hydrocarbons and the model uncertainties caused by chemical rates therefore more limited. In the case of Neptune, \citet{Dobrijevic2020} showed that uncertainties on model results for \ce{H2O} are very low compared to other oxygen species and hydrocarbons and this species is therefore currently the best tracer of the vertical diffusion in the stratosphere of Neptune, assuming its chemistry is well-known. We considered in the present paper that this was also the case for Jupiter (and the other giant planets). 

The delivery of \ce{H2O}, among other species like HCN, CO and carbon sulfide (CS), to Jupiter's stratosphere by comet SL9 in 1994 \citep{Lellouch1995} further enhances the interest of using these species as tracers for horizontal and vertical dynamics in this atmosphere, provided that either they are chemically stable over the time considered or their chemistry is properly modeled. For instance, \citet{Moreno2003}, \citet{Griffith2004} and \citet{Lellouch2006} used HCN, CO and \ce{CO2} to constrain longitudinal and mostly meridional diffusion in the years following the impacts, even though \ce{CO2} is not chemically stable \citep{Lellouch2002}. In the present study, we used nearly two decades of \ce{H2O} disk-averaged emission monitoring with Odin and a 1D photochemical model to constrain vertical diffusion in Jupiter's stratosphere. Not only does the modeling of the \ce{H2O} vertical profile suffer less from chemical rate uncertainties than hydrocarbons \citep{Dobrijevic2020}, but the progressive downward diffusion of \ce{H2O} from its initial deposition level ($p_0$ in our model) enabled us to probe $K_{zz}$ at various altitudes as a function of time. This will remain true for the years to come, until the bulk of \ce{H2O} eventually reaches its condensation level at $\sim$30\,mbar. 

When assuming Model A for $K_{zz}$, we cannot fit the $\sim$40\%~decrease observed on the l/c, even if we find a global decrease of the \ce{H2O} column of 5\% between the impacts and 2019. \ce{H2O} is too efficiently recycled for its profiles to reproduce the time series of Odin observations. We can only fit this time series with the 1D model if we alter $K_{zz}$ to Model B. While the initial loss of \ce{H2O} is caused by the build-up of the \ce{CO2} column, condensation becomes the main loss factor after $\sim$2$\times$10$^8$\,s (about 6 years after the impacts) and enables fitting the \ce{H2O} observations. \ce{CO2} also starts being lost to \ce{H2O} and subsequent condensation of \ce{H2O}.

While the Odin time series of \ce{H2O} observations can be fitted with our 1D model and $K_{zz}$ Model B, the resulting C$_2$H$_x$ profiles are inconsistent with numerous observations, even when accounting for the joint error bar of the observations and the photochemical model itself. This tends to demonstrate that our 1D model cannot fit jointly C$_2$H$_x$ and the \ce{H2O} observation time series. 
$K_{zz}$ Model A probably remains the best disk-averaged estimate of $K_{zz}$ in Jupiter's stratosphere. In this context, Model B can only be seen as an upper limit. \ce{H2O} must then have an additional loss process. 

We propose as a promising candidate that the regions of enhanced loss of \ce{H2O} are the auroral regions of Jupiter, by means of ion-neutral chemistry. \citet{Dobrijevic2020} showed that ion-neutral chemistry affected the abundances of oxygen species in Neptune's atmosphere, even without including magnetospheric ions and electrons. With energetic electrons precipitating down to the submillibar level under Jupiter's aurorae \citep{Gerard2014}, ion-neutral chemistry could be the cause for an enhanced loss of \ce{H2O}, possibly producing excess \ce{CO2}. This could explain the peak in the \ce{CO2} meridional distribution seen 6 years after the SL9 impacts only at the south pole by \citet{Lellouch2006}, as SL9-originating CO and \ce{H2O} had not yet reached the northern polar region \citep{Moreno2003,Cavalie2013}. It must be noted that unexpected distributions of hydrocarbons have already been found in Jupiter's auroral regions. \citet{Kunde2004} and \citet{Nixon2007} found that, contrary to \ce{C2H2}, the zonal mean of \ce{C2H6} did not follow the mean insolation and peaked at polar latitudes. \citet{Hue2018} demonstrated that this discrepancy between two species that share a similar neutral chemistry cannot be explained either by neutral chemistry or by a combination of advective and diffusive transport. In turn, they proposed ion-neutral chemistry in the auroral region as a mechanism to bring the zonal  mean of \ce{C2H6} out equilibrium with the solar insolation. More recently, \citet{Sinclair2018} measured the longitudinal variability of the main C$_2$H$_x$ species at northern and southern auroral latitudes. They found that \ce{C2H2} and \ce{C2H4} were significantly enhanced at millibar and submillibar pressures under the aurora, while \ce{C2H6} remained fairly constant. In addition, \citet{Sinclair2019} found that heavier hydrocarbons like \ce{C6H6}were also enhanced under the aurorae. This points to a richer chemistry than that seen at lower latitudes, increasing the production of several hydrocarbons and ultimately producing Jupiter's aerosols \citep{Zhang2013,Zhang2016,Giles2019}. Such a richer chemistry could also apply to H$_2$O and other oxygen species. At this point, however, this remains speculative and requires modeling the auroral chemistry under Jovian conditions with and without SL9 material.


\section{Conclusion} 
\label{section:conclusion}

In this paper, we presented disk-averaged observations of \ce{H2O} vapor in the stratosphere of Jupiter carried out with the Odin space telescope. This temporal monitoring of the \ce{H2O} line at 557 GHz spans over nearly two decades, starting in 2002, i.e. 8 years after its delivery by comet SL9. We demonstrated that the line-to-continuum ratio has been decreasing as a function of time by $\sim$40\%~on this period. Such a trend results from the evolution of the \ce{H2O} disk-averaged vertical profile and we used it to study the chemistry and dynamics of the Jovian atmosphere. 

We thus used our observations to constrain $K_{zz}$ in the levels where \ce{H2O} resided at the time of our observations, i.e. between $\sim$0.2 and $\sim$5\,mbar. Using a combination of photochemical and radiative transfer modeling, we showed that the $K_{zz}$ profile of \citet{Moses2005} could not reproduce the observations. We had to increase the magnitude of $K_{zz}$ by a factor of $\sim$2 at 0.1\,mbar and $\sim$4 at 1\,mbar to fit the full set of Odin observations. 

However, this $K_{zz}$ profile makes the \ce{C2H6} profile fall outside observational and photochemical model error bars and is thus not acceptable. As a result, 1D time-dependent photochemical models cannot reproduce both the main hydrocarbon profiles and the temporal evolution of the disk-averaged \ce{H2O} vertical profile. A possible explanation is that these species still vary locally more sharply as a function of latitude than the factor of 2-3 indicated by the low spatial resolution observations of \citet{Cavalie2013} (these variations cannot be studied with 1D models, by definition). \citet{Sinclair2018,Sinclair2019} have already demonstrated that the auroral regions of Jupiter harbor chemistry influencing the hydrocarbons that is not seen elsewhere on the planet. The same may be also be true for \ce{H2O}, but disk-resolved observations with more resolution that did Herschel in 2009-2010 would be required to test this hypothesis, possibly with the James Webb Space Telescope \citet{Norwood2016}. In the meantime, the continuation of the monitoring of the Jovian stratospheric \ce{H2O} emission with Odin will help prepare future observations to be carried out by the Jupiter Icy Moons Explorer (JUICE).

The study that we have presented in this paper will help to prepare the JUICE mission which will study the Jupiter and its moons in the 2030s. One instrument of its payload, the Submillimetre Wave Instrument (SWI; \citealt{Hartogh2013}) will observe the same \ce{H2O} line as the one observed by Odin to map the zonal winds in the stratosphere of Jupiter from high resolution spectroscopy at high spatial resolution. The continuation of the Odin monitoring is thus crucial to refine our estimates of the \ce{H2O} abundance and vertical profile for the 2030s and thus optimize the SWI observation program.

\begin{acknowledgements}
This work was supported by the Programme National de Plan\'etologie (PNP) of CNRS/INSU and by CNES. Odin is a Swedish-led satellite project funded jointly by the Swedish National Space Board (SNSB), the Canadian Space Agency (CSA), the National Technology Agency of Finland (Tekes), the Centre National d'\'Etudes Spatiales (CNES), France and the European Space Agency (ESA). The former Space division of the Swedish Space Corporation, today OHB Sweden, is the prime contractor, also responsible for Odin operations.
\end{acknowledgements}

\bibliographystyle{aa} 

\end{document}